\documentclass[aps,prd,twocolumn,superscriptaddress,showpacs,showkeys,amsfonts,amssymb,amsmath]{revtex4-1}

\usepackage{graphicx}
\usepackage[colorlinks=true,linkcolor=blue,citecolor=blue,urlcolor=blue]{hyperref}

\newcommand{\co}{(Color online) }

\begin{document}

\title{New Exclusion Limits for the Search of Scalar and Pseudoscalar Axion-Like Particles from \textquotedblleft~Light Shining Through a Wall~\textquotedblright}

\author{R.~Ballou}
\affiliation{CNRS, Institut N\'eel, 38042 Grenoble, France}
\affiliation{Universit\'e Grenoble Alpes, 38042 Grenoble, France}
\author{G.~Deferne}
\affiliation{CERN, CH-1211 Geneva-23, Switzerland}
\author{M.~Finger~Jr.}
\affiliation{Charles University, Faculty of Mathematics and Physics, Prague, Czech Republic}
\author{M.~Finger}
\affiliation{Charles University, Faculty of Mathematics and Physics, Prague, Czech Republic}
\author{L.~Flekova}
\affiliation{Czech Technical University, Prague, Czech Republic}
\author{J.~Hosek}
\affiliation{Czech Technical University, Prague, Czech Republic}
\author{S.~Kunc}
\affiliation{Technical University of Liberec, 46117 Liberec, Czech Republic}
\author{K.~Macuchova}
\affiliation{Czech Technical University, Prague, Czech Republic}
\author{K.~A.~Meissner}
\affiliation{University of Warsaw, Institute of Theoretical Physics, 00-681 Warsaw, Poland}
\author{P.~Pugnat}
\email{pierre.pugnat@lncmi.cnrs.fr}
\affiliation{CNRS, LNCMI, F-38042 Grenoble, France}
\affiliation{Universit\'e Grenoble Alpes, 38042 Grenoble, France}
\author{M.~Schott}
\affiliation{University of Mainz, Institute of Physics, 55128 Mainz, Germany}
\author{A.~Siemko}
\affiliation{CERN, CH-1211 Geneva-23, Switzerland}
\author{M.~Slunecka}
\affiliation{Charles University, Faculty of Mathematics and Physics, Prague, Czech Republic}
\author{M.~Sulc}
\affiliation{Technical University of Liberec, 46117 Liberec, Czech Republic}
\author{C.~Weinsheimer}
\affiliation{University of Mainz, Institute of Physics, 55128 Mainz, Germany}
\author{J.~Zicha}
\affiliation{Czech Technical University, Prague, Czech Republic}

\collaboration{OSQAR Collaboration}

\date{\today}

\begin{abstract}
{
Physics beyond the Standard Model predicts the possible existence of new particles that can be searched at the low energy frontier in the sub-eV range. The OSQAR photon regeneration experiment looks for \textquotedblleft Light Shining through a Wall\textquotedblright~ from the quantum oscillation of optical photons into \textquotedblleft Weakly Interacting Sub-eV Particles\textquotedblright, such as axion or Axion-Like Particles~(ALPs), in a 9\,T transverse magnetic field over the unprecedented length of $2 \times 14.3$\,m. In 2014, this experiment has been run with an outstanding sensitivity, using an 18.5\,W continuous wave laser emitting in the green at the single wavelength of 532\,nm. No regenerated photons have been detected after the wall, pushing the limits for the existence of axions and ALPs down to an unprecedented level for such a type of laboratory experiment. The di-photon couplings of possible pseudoscalar and scalar ALPs can be constrained in the nearly massless limit to be less than $3.5\cdot 10^{-8}$\,GeV$^{-1}$ and $3.2\cdot 10^{-8}$\,GeV$^{-1}$, respectively, at 95\% Confidence Level.
}
\end{abstract}

\pacs{14.80.Va, 14.80.-j, 95.35.+d, 14.70.Bh}
\keywords{Axion, Scalar and Pseudoscalar Axion-like Particles, Photon Regeneration}

\maketitle

\section{Introduction}
\label{sec:introduction}

Possible extensions of the Standard Model (SM) of particle physics are not restricted to the high energy frontier. There is also a growing interest for the search of Weakly Interacting Sub-eV Particles (WISPs), with much weaker interactions and masses below the eV. One emblematic example is the axion, a pseudoscalar boson arising from the spontaneous breaking of a global chiral symmetry $U(1)_A$, postulated to dynamically solve the strong CP problem~\cite{Peccei1977, Weinberg1977, Wilczek1977}. Axions and Axion-Like Particles (ALPs) are predicted in supersymmetric theories \cite{Covi1999}, in string theory \cite{Svrcek2006, Cicoli2012}, and in the Conformal Standard Model \cite{Meissner2007}. ALPs can be scalar as well as pseudoscalar and theorized to couple to the SM through a variety of mechanisms, giving rise in particular to a two-photons vertex. WISPs also include light bosons of gauge groups under which the SM particles are not charged (hidden sectors), which may interact with the SM through gravity, kinetic mixing or higher order quantum processes \cite{Holdom1986}. The interest aroused by WISPs goes beyond particle physics. As earlier hypothesized for the axion \cite{Abbott1982}, they provide alternative candidates for Dark Matter (DM) \cite{Bradlay2003, Arias2012, Ringwald2012}. Moreover, they might explain a number of astrophysical puzzles, such as the universe transparency to very high energy photons ($>$ 100\,GeV) \cite{Meyer2013}, the anomalous white dwarf cooling \cite{Bertolami2014} or the recently discovered gamma ray excesses in galaxy clusters \cite{Cicoli2014}.
In contrast to Weakly Interacting Massive Particles (WIMPs), which can be searched for at TeV colliders such as the Large Hadron Collider (LHC) at CERN, the detection of WISPs requires to have recourse to dedicated low-energy experiments. Several methodologies exploiting the existence of a di-photon coupling have been proposed based on lasers, microwave cavities, strong electromagnetic fields or torsion balances \cite{JaRi2010, AWGr2011}. 

The OSQAR (Optical Search for QED Vacuum Birefringence, Axions and Photon Regeneration) experiment at CERN, is at the forefront of this low-energy frontier of particle/astroparticle physics. It combines the simultaneous use of high magnetic fields with laser beams in distinct experiments. One of its setups uses the ``Light Shining Through a Wall'' (LSW) method for the search of the ALPs \cite{Bibber1987}. A pioneering work in this line excluded ALPs with a di-photon coupling constant $g_{\text{A}\gamma\gamma}$ larger than $6.7 \cdot 10^{-7}$\,GeV$^{-1}$ for masses below $10^{-3}$\,eV~\cite{Cameron1993}. These exclusion limits were later extended by other LSW experiments to $g_{\text{A}\gamma\gamma} > 6.5 \cdot 10^{-8}$\,GeV$^{-1}$ for masses below $5\times 10^{-4}$\,eV~\cite{Ehret2010} and more recently tightened to $g_{\text{A}\gamma\gamma} > 5.7 \cdot 10^{-8}$\,GeV$^{-1}$ for masses below $2 \times 10^{-4}$\,eV~\cite{Pugnat2014}. 
New results are here reported, obtained from the 2014 data-taking campaign of the OSQAR LSW experiment. Compared to the previous experimental setups of OSQAR \cite{Pugnat2014, Pugnat2013}, a more powerful laser source and a detector with higher sensitivity have been used. In addition, an improved data analysis strategy has been implemented.

\section{Experimental Setup and Data Taking}
\label{sec:setup}

\begin{figure}[t]
    \centering
    \includegraphics[width=0.48\textwidth]{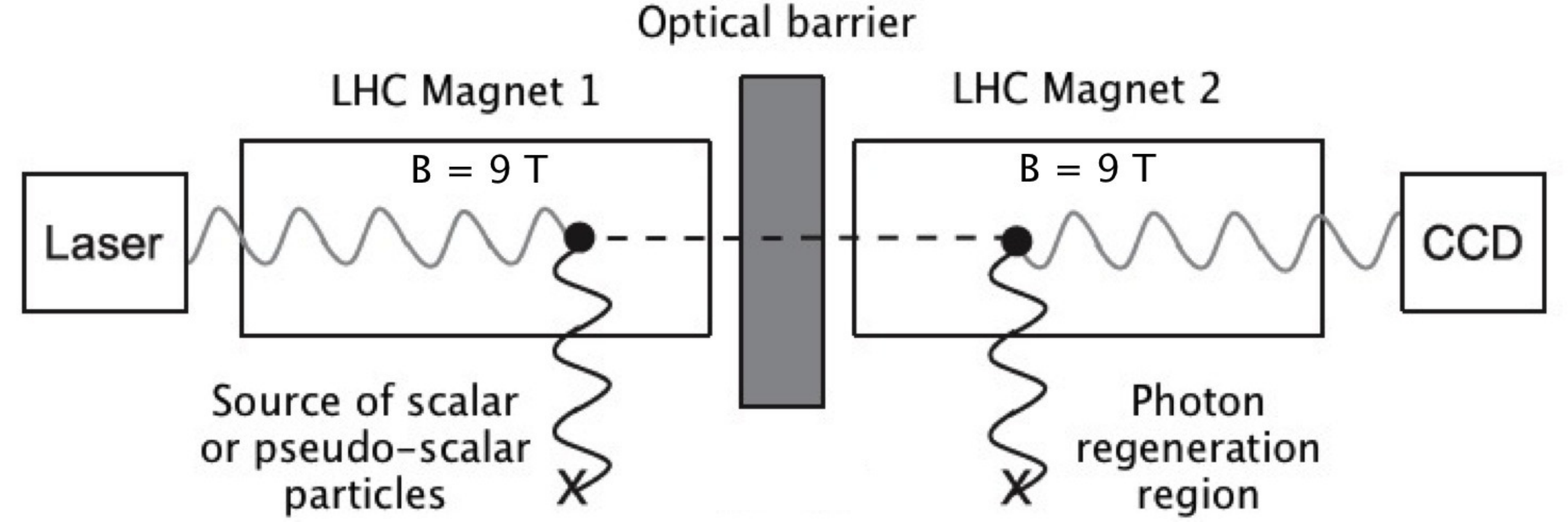}
\caption{Principle of the OSQAR LSW experiment.}
\label{fig:lsw}
\end{figure}

LSW experiments are based on the combination of two factors. The first is the transparency of the WISPs to photons barriers, owing to the weakness of the interactions of the WISPs with the particles of the SM. The second is the photon-to-WISP and WISP-to-photon quantum oscillation, which would arise from the interactions of WISPs with photons~\cite{Sikivie1983, Bibber1987, Arias2010}.  
In the case of the sub-eV ALPs, the method benefits from their two-photon vertex, inducing oscillations with optical photons in a transverse magnetic field. Such a field can be represented as a sea of virtual photons, whose interaction with real photons (resp. ALPs) can produce real ALPs (resp. photons). The mechanism is similar to the Primakov process of the production of neutral mesons by high-energy photons in a strong electric field \cite{Primakov1951}. 
The effective Lagrangian density of the interaction of an axion or a pseudoscalar ALP (PS-ALP) field $\mathcal{A}$ with the electromagnetic field $F_{\mu\nu}$ is written generically in the form 
\begin{equation}
\mathcal{L}_{int} = -\frac{1}{4} g_{\text{A}\gamma\gamma}~\mathcal{A}~F_{\mu\nu}\widetilde{F}^{\mu\nu} = g_{\text{A}\gamma\gamma}~\mathcal{A}~\vec{E} \cdot \vec{B}
    \label{eq:effectivepsalp}
\end{equation}
where $\widetilde{F}^{\mu\nu} = \frac{1}{2}\epsilon^{\mu\nu\alpha\beta}F_{\alpha\beta}$ is the dual of $F_{\mu\nu}$ and the constant $g_{\text{A}\gamma\gamma}$ stands for the effective axion or ALP di-photon coupling. 
With a Scalar ALP (S-ALP) field $\mathcal{A}$ the interaction with the field $F_{\mu\nu}$ takes the generic form 
\begin{equation}
\mathcal{L}_{int} = -\frac{1}{4} g_{\text{A}\gamma\gamma}~\mathcal{A}~F_{\mu\nu} F^{\mu\nu} = g_{\text{A}\gamma\gamma}~\mathcal{A}~\frac{1}{2}(\vec{E}^2 - \vec{B}^2)
    \label{eq:effectivesalp}
\end{equation}
Accordingly, either PS-ALPs or S-ALPs could potentially be created when a beam of linearly polarized photons propagates in a transverse magnetic field $\vec{B}$, depending on whether the polarization is parallel to the magnetic field ($\vec{E}_\gamma \parallel \vec{B}$) or perpendicular ($\vec{E}_\gamma \perp \vec{B}$). 
If an optical barrier is placed downstream to the beam, all unconverted photons will be absorbed while ALPs would traverse the optical barrier. By applying a second magnetic field in the regeneration domain beyond the wall, the inverse Primakov process can convert the ALPs back into photons, which can be subsequently detected (Fig.~\ref{fig:lsw}). 
The probability of an ALP-to-photon ($A \rightarrow \gamma$) or of a photon-to-ALP ($\gamma \rightarrow A$) conversion is given by~\cite{Sikivie1983, Bibber1987, Arias2010}:
\begin{equation}
P_{\gamma\leftrightarrow A} = \frac{1}{4} {(g_{\text{A}\gamma\gamma} BL)}^2
        {\left(
            \frac{2}{qL} \sin \frac{qL}{2}
        \right)}^2
    \label{eq:probability}
\end{equation}
in units of Heaviside-Lorentz system $(\hbar=c=1)$. $q = | k_\gamma - k_A |$ stands for the momentum transfer, where $k_\gamma = \omega$ is the momentum of the photon of energy $\omega$ and $k_A =
\sqrt{(\omega^2 - m_A^2)}$ the momentum of the ALP of mass $m_A$.
The overall probability of the photon regeneration is $P_{\gamma \rightarrow A \rightarrow \gamma} = {(P_{\gamma\leftrightarrow A})}^2$, as arising from the two consecutive conversions: $\gamma \rightarrow A$ followed by $A \rightarrow \gamma$. The flux of the regenerated photons to detect is then given by 
\begin{equation}
    \frac{dN}{dt} = \frac{P}{\omega} \eta \; {(P_{\gamma\leftrightarrow A})}^2
    \label{eq:flux}
\end{equation}
where $P$ represents the power of the incoming photon beam and $\eta$ stands for the photon detection efficiency. $dN/dt$ is proportional to the $4$-th power of the field integral $BL$, whence the need for strongest magnetic field $B$ over longest optical path $L$ to achieve highest sensitivity. 

The OSQAR LSW experiment uses two spare LHC dipole magnets, the first for the ALPs production and the second for the photon regeneration. Each magnet is cooled down to 1.9\,K with superfluid He and provides an uniform transverse magnetic field with a strength of 9\,T over a magnetic length of 14.3\,m, thus giving rise to a magnetic field integral of $BL = 128.7$\,Tm. The aperture of both magnets have been pumped, using turbo-molecular pumping groups, down to $10^{-5}-10^{-7}$\,mbar.

A continuous wave optical power of 18.5\,W at a single wavelength of 532\,nm (2.33\,eV) was delivered by a diode-pumped solid-state laser (Verdi V18 from COHERENT Inc.). A beam expander telescope has been used to reduce the laser beam divergence. The photon beam is linearly polarized with a vertical orientation parallel to the magnetic field for the search of PS-ALPs. A $\lambda/2$ wave-plate with antireflective (AR) coating layers was inserted at the laser exit in order to align the photon polarization in the perpendicular direction for the search of S-ALPs. 

The laser beam at the exit of the second magnet aperture was focused by an optical lens on a thermoelectric cooled Charge-Coupled Device (CCD) with an AR coated window (DU934P-BEX2-DD from ANDOR Co.). The CCD chip has an active area of $13.3 \times 13.3$\,mm$^2$ and is composed by a 2D array of $1024 \times 1024$ square pixels of size $13 \times 13$ $\mu$m$^2$. The laser beam was focussed at $95 \%$ on an area covering not more than 4 pixels.
The CCD was cooled down to the temperature range [$-95^\circ$\,C, $-92^\circ$\,C]. The dark current is 0.0012\,e$^-$/pixel/s at $-95^\circ$\,C and 0.0020\,e$^-$/pixel/s at $-92^\circ$\,C. The readout noise was determined at 2.5\,e$^-$rms/pixel from frames recorded in minimal time with closed shutters. These measured CCD parameters are in agreement with those provided by the manufacturer.
The overall photon detection efficiency was explicitly measured to $\eta = 0.56~\pm~0.02$\,ADU/photon, taking into account the losses coming from optical elements including windows, plates, lens, CCD quantum efficiency of 0.88 at 532\,nm and sensitivity of 1.3\,e$^-$/Analog-to-Digital Unit (ADU). The total optical power loss of the whole optical setup is thus approximately $17\%$, in good agreement with the characteristics of the various optical elements in use.

The data taking for the ALPs search reported in this paper was performed in August 2014 and corresponds to a total of 119 experimental runs, as detailed in Table \ref{exppar}. 

\begin{table}[h]
\centering
\begin{tabular*}{\columnwidth}{@{\extracolsep{\fill}}|c|c|c|c|c|c|c|@{}}
\hline
Search for & $\vec{E}_\gamma$ & $\vec{B}$  & $P$  & $N_T$ & $N_R$ & W(95\%) \\ 
\hline PS-ALPs & $\parallel \vec{B}$ & 9\,T & 18.5\,W & 59 & 18 & 0.64\,mHz \\ 
\hline S-ALPs & $\perp \vec{B}$ & 9\,T & 18.5\,W & 60 & 12 & 0.45\,mHz \\ 
\hline
\end{tabular*}
\caption{Summary of the experimental runs. $\vec{E}_\gamma$ stands for the electric component of the linearly polarized photon field, $\vec{B}$ the static magnetic field, $P$ the laser power, $N_T$ the number of recorded runs and $N_R$ the number of rejected runs. W(95\%) is the Bayesian threshold of non detection at 95\%.}
\label{exppar}
\end{table}

\noindent Each experimental run was composed of two frames of 5400\,s exposure time separated by a 100\,s pause. The recording time for a single frame was defined to optimize the signal over noise ratio, taking into account the filtering and removal procedure of cosmic ray signatures. Before and after each run, the position of the laser beam was precisely measured on the CCD with the optical power strongly reduced. Starting with a laser beam power settled to 3\,W, a variable beam splitter has been introduced to attenuate the beam typically by a factor of 1/500. To carefully check the stability of the laser beam spot, three frames were recorded each time with 0.01\,s exposure separated by a 120\,s pause. Subsequently the optical barrier was introduced and the attenuator was settled to its minimum before ramping up the laser power up to 18.5\,W. If the experimental conditions were not optimal or if cosmic rays have impacted the signal region then the experimental run was rejected (Table \ref{exppar}). The signal region is defined as the pixels of the CCD where reconverted photons are expected to be detected, taking into account slight displacements of the laser beam spot. 
%

\section{Data Reduction}
\label{sec:datareduction}

\begin{figure}[t]
    \centering
    \includegraphics[width=0.45\textwidth]{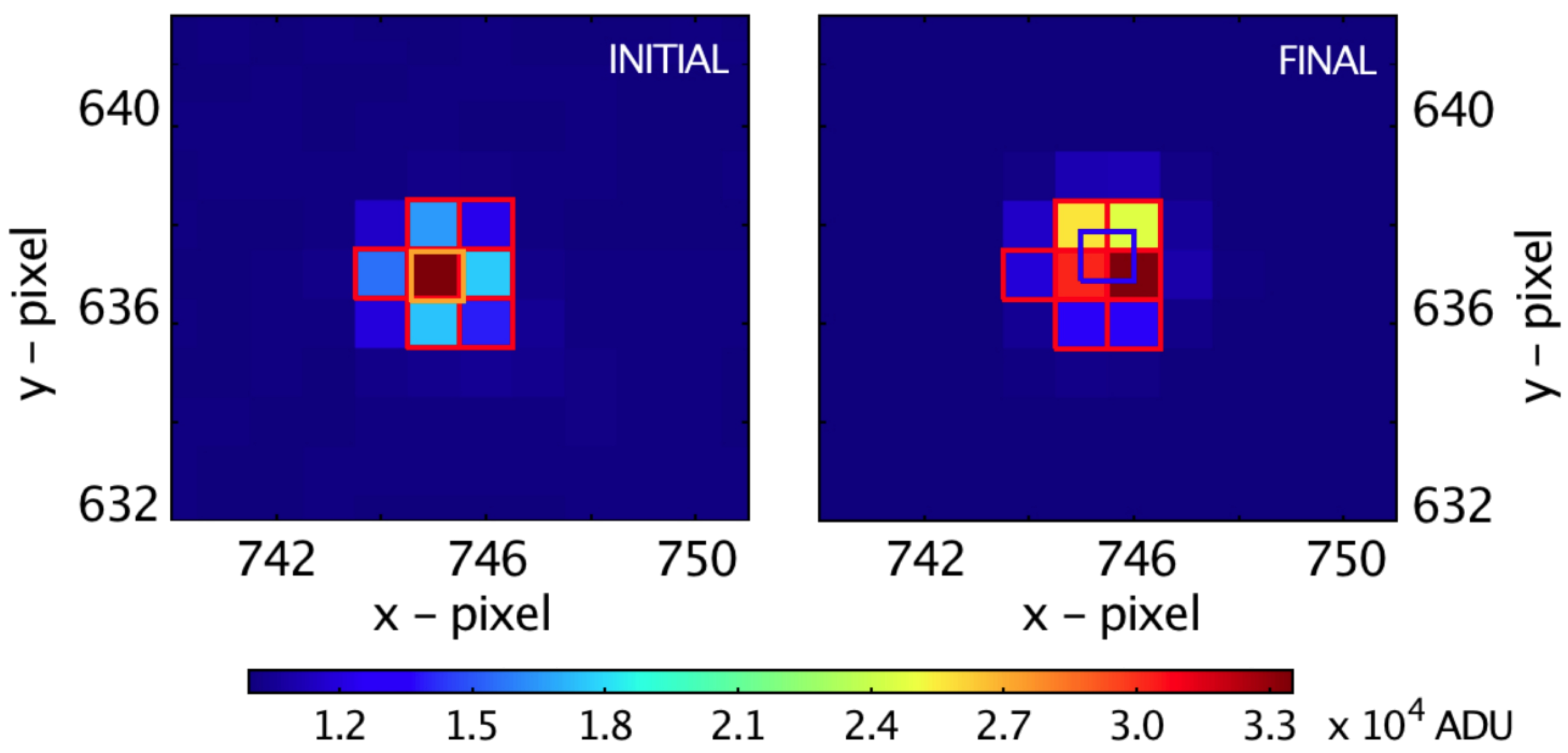}
\caption{\co Initial and final positions of the attenuated laser beam on the CCD for Run-90 devoted to the search of S-ALPs ($\vec{E}_\gamma \perp \vec{B}$). The signal region covers 7 pixels.}
\label{fig:sigreg}
\end{figure}

The detailed analysis of all beam positions during the full data taking period clearly confirmed
that environmental temperature variations induce only minor low-frequency displacement/deformation of the CCD support and therefore only a small relative shift of the beam spot on the detector, typically less or equal to one pixel per hour. The beam spot displacement is unidirectional, with no oscillation observed after the exposure time of a full run, \textit{i.e.} after $2 \times 5400\,\text{s} + 100\,\text{s} = 10900$\,s. The beam spot sizes before and after each experimental run are obtained from least squares fits with a two-dimensional Gaussian distribution. To define conservatively the signal region, twice the maximum value of the gaussian widths computed from the \textit{intitial} (\textit{i}) and \textit{final} (\textit{f}) reference frames is retained: $2\, \sigma_\text{Run} = 2 \times \text{max}\bigl[\sigma_i, \sigma_f \bigr]$. If the laser positions before and after a run do not coincide, all pixels within a $2\, \sigma_\text{Run}$ orthogonal distance along a straight line between the final and initial positions are added to the signal region. An illustrative example is provided in Fig.~\ref{fig:sigreg}. It follows that the number $S$ of pixels that defines the signal region can vary from run to run (Table \ref{sigregsize}). 
A cumulated count $N_i$ can be associated to each Run-i, by adding the counts collected during the exposure time of $2 \times 5400\,\text{s}$ for ALPs detection on all the pixels of the signal region, once this one is carefully determined. $N_i$ is the relevant quantity that would reveal the regenerated photons if it is proved in excess over background noises.
\begin{table}[h]
\centering
\begin{tabular*}{\columnwidth}{@{\extracolsep{\fill}}|c|c|c|c|c|c|c|c|c|c|c|c|c|c|c|c|@{}}
\hline
$S$ & ~6  & ~7  & ~8  & ~9  & 10 & 11 & 12 & 13 & 14 & 15 & 16 & 17 & 18 & 19 & 20 \\ 
\hline $N_{PS}$ & ~0 & ~2 & ~6 & ~4 & ~3 & ~4 & ~7 & ~4 & ~3 & ~2 & ~0 & ~2 & ~3 & ~0 & ~1 \\ 
\hline $N_{S}$ & ~6 & ~5 & ~6 & 11 & ~3 & ~4 & ~5 & ~3 & ~2 & ~1 & ~1 & ~0 & ~0 & ~0 & ~1 \\ 
\hline
\end{tabular*}
\caption{Distribution of sizes $S$ in unit of pixel of the signal regions. $N_{PS}$ (resp. $N_{S}$) counts the number of runs devoted to the search of PS-ALPs (resp. S-ALPs) for which the same given size $S$ was experimentally deduced for the signal region.}
\label{sigregsize}
\end{table}
\begin{figure}[b]
    \centering
    \includegraphics[width=0.45\textwidth]{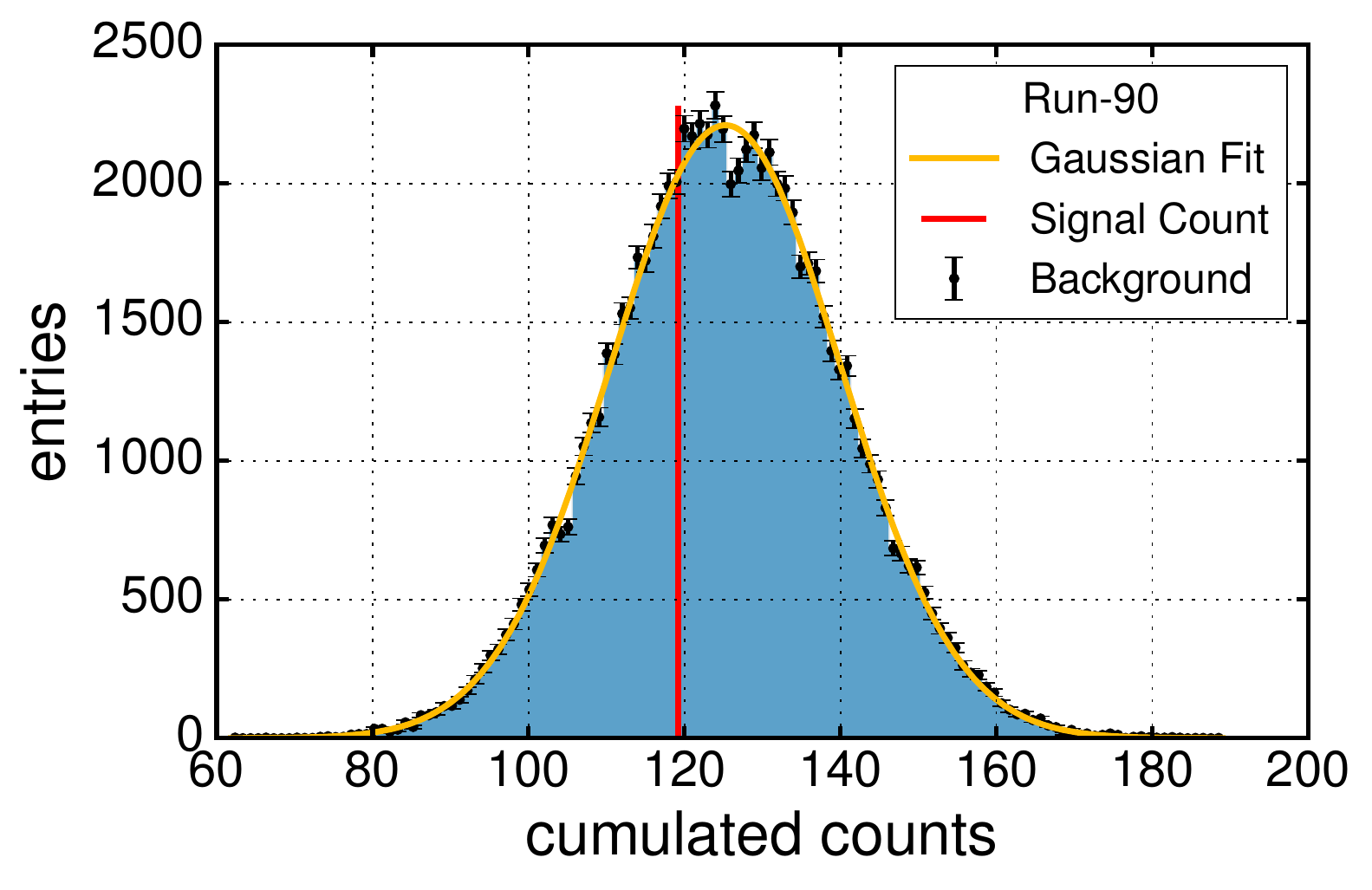}
\caption{ \co Distribution of the cumulated counts $n_{90j}$ of background clusters $j$ of size and shape identical to the signal region shown Fig.~\ref{fig:sigreg} for Run-90, after applying all background corrections (see text). Gaussian fit gives mean $\mu_{90}^{\text{bkg}} = 125.24 \pm 0.05$ and standard deviation $\sigma_{90}^{\text{bkg}} = 14.77 \pm 0.04$ with reduced chi-squared $\chi^2 / n.d.f = 1.04$. The red line represents the cumulated count $N_{90}=119$ of the signal region. It crosses the Gaussian distribution at about 2000, which corresponds to the number of background clusters with the same cumulated count as the signal region.}
\label{fig:countc}
\end{figure}

One of the advantages of a small and well-defined signal region is the possibility to use the remaining pixels, which are not exposed to possible ALPs signals, to fully characterize the background without repeating dedicated background acquisition runs \cite{Pugnat2014, Pugnat2013}. 
These remaining pixels are clustered in size and shape that correspond to the signal region. 
Cumulated counts $n_{ij}$ are obtained for each cluster $j$ in the background region of each Run-i, by adding the counts collected on all the pixels of the chosen cluster, similarly to the cumulated count $N_i$ of the signal region. The distribution of the $n_{ij}$ for a given Run-i then provides an estimate of the expected background contribution to $N_i$.
As it is well known for CCDs, the width of this distribution arises mainly from two independent sources, namely dark-counts that increases with exposure times and read-out noise that arises from the analog-digital conversion process of the readout-system.

Before analyzing the cumulative counts, three corrections have been applied to each recorded frame pair of $2 \times 5400$\,s. 
The first one aims to correct long wavelength distortions of the spatial distribution of the background arising from the minor heterogenity in the cooling of the CCD chip. Corrections of these background drifts are achieved by median filtering.
The objective of the second correction is to reduce the so-called \textit{fixed pattern noise} evolving from possibly unequal pixel biases. As this noise manifests only during readout, and is independent of the exposure time, it has been estimated by capturing dark frames (\textit{i.e.} shutter closed) with the minimal exposure time (10$^{-5}$\,s in our case). Ten frames have been recorded under these conditions and have been averaged pixel-wise to define a \textit{bias frame}, which is subtracted from every frame used in the data analyses. This correction also subtracts the general constant offset of $\approx 3300$ counts on each pixel imprinted by the chip in each frame of 5400\,s. 
The last correction step removes those pixels which were hit by cosmic rays, using a short wavelength filtering process. Whenever a pixel count exceeds the significant threshold of the distribution of pixel counts in the background region this pixel is rejected together with its nearest neighbours. In case the signal region is hit by a cosmic ray, the overall frame is rejected.

Once all correction steps have been applied to a given Run-i, the cumulated counts $N_i$ in the signal region and $n_{ij}$ of the clusters $j$ in the background region are evaluated. 
As illustrated in Fig.~\ref{fig:countc} for the Run-90, the distribution of the cumulated counts $n_{ij}$ for every Run-i is a gaussian with mean $\mu_{i}^{\text{bkg}}$ and standard deviation $\sigma_{i}^{\text{bkg}}$ in good agreement with the total noise of the CCD expected from dark-counts and read-out noise contributions.

\section{Bayesian Analysis}
\label{sec:analysis}

A detection limit on the rate of reconverted photons $dN/dt$ can be extracted from the gaussian width of cumulated background cluster counts over the pixel-wise sum of all the frames for a given photon polarization. This requires choosing a signal region of the same size for all the runs, which necessarily must be the largest one that is $S=20$ pixels according to Table \ref{sigregsize}. 
A slightly better detection limit can be obtained by rejecting the few runs with too large size of signal region, but at the statistical cost of the counts inherent to the corresponding decrease of the total exposure time. This strategy used in previously published analyses of similar experiments, e.g. Ref. \cite{Pugnat2013}, is not optimal. A detection limit at two times the standard deviation of the background fluctuations in the order of 2\,mHz ($2 \times 10^{-3}$\,ADU\,s$^{-1}$) can be obtained with this approach.  This does not constitute a real  improvement with respect to a single run with a smaller size of  signal region such as Run-90 for which $S=7$ pixels, because the increase in cluster size leads to larger background fluctuations for such a run. 
\begin{figure}[b]
\begin{tabular}{cc}
   \includegraphics[width=0.24\textwidth]{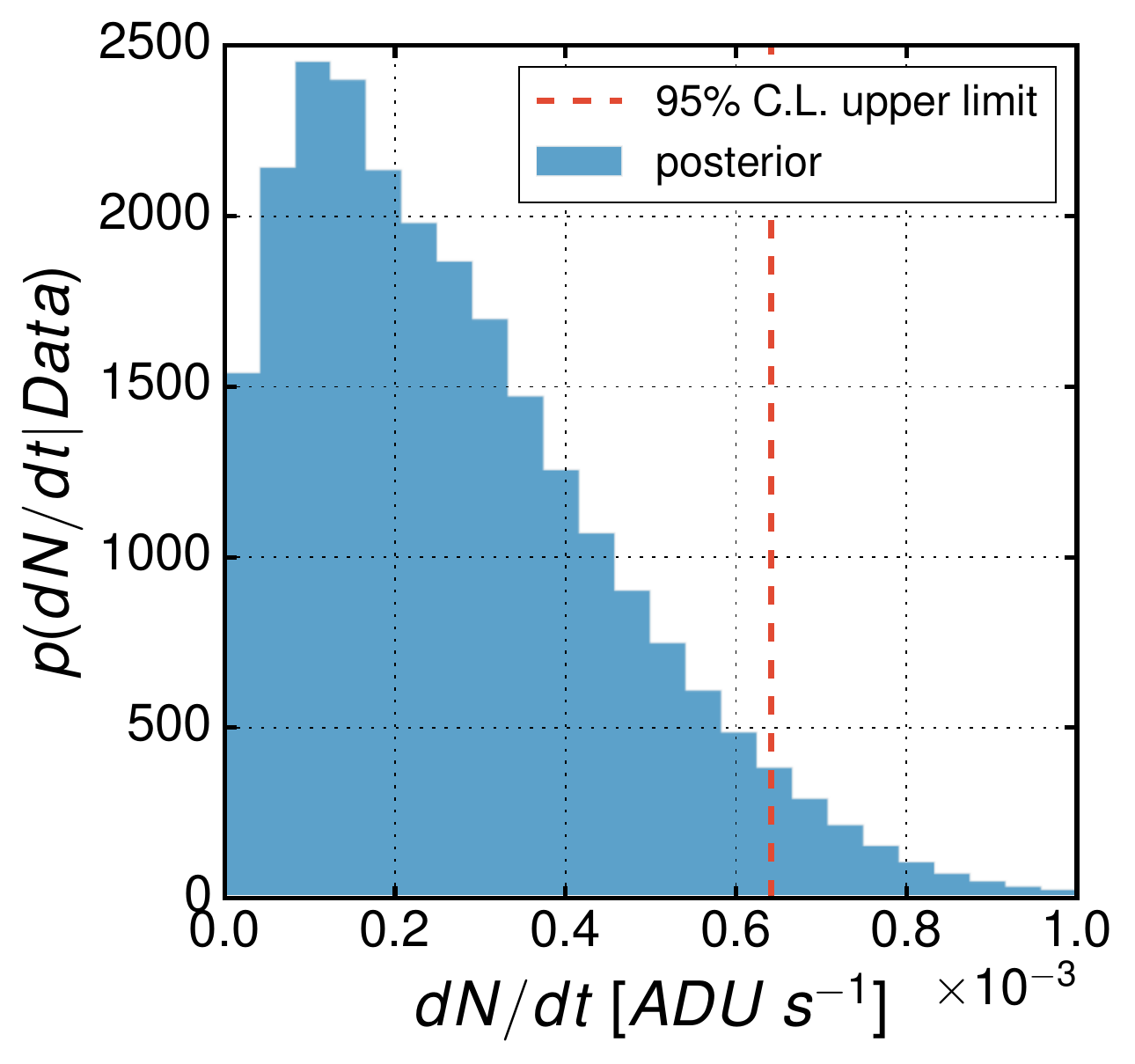}
 &
   \includegraphics[width=0.24\textwidth]{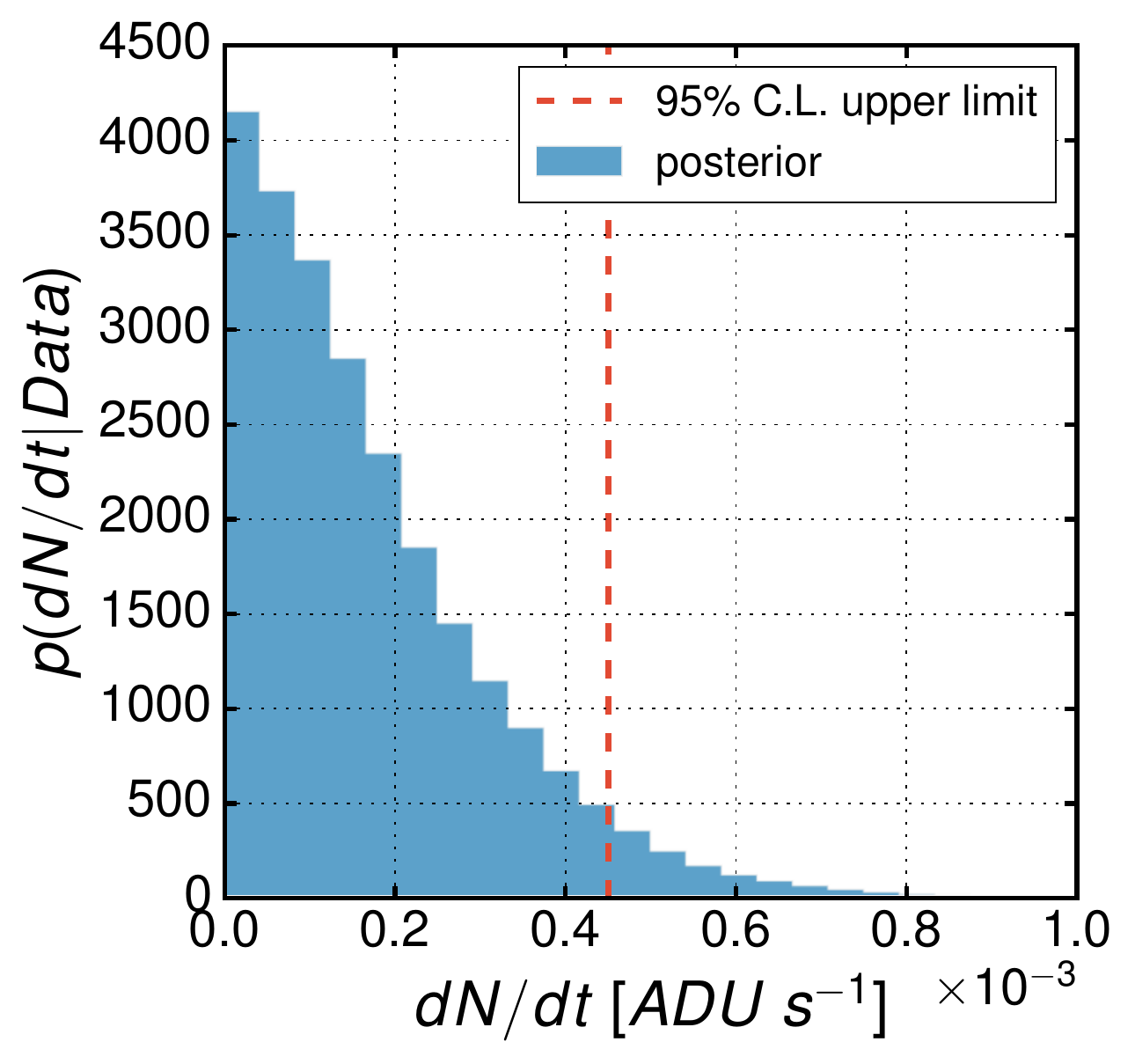}
\end{tabular}
    \caption{ \co Posterior probability distribution function  $p(dN/dt ~|~ Data)$ of reconverted photon rate $dN/dt$ for incoming photon polarization parallel ($\vec{E}_\gamma \parallel \vec{B}$) (Left) and perpendicular ($\vec{E}_\gamma \perp \vec{B}$) (Right) to the magnetic field $\vec{B}$.}
    \label{fig:postdistr}
\end{figure}

In order to optimize the detection limit by considering real signal regions of different sizes from run to run (Table \ref{sigregsize}), a Bayesian approach has been implemented (cf. Ref. \cite{Olive2014} pages 472-487). The following Likelihood model has been considered  
\begin{equation}
    \mathcal{L} \propto
    \prod_{i} \mathcal{N}\Bigl(N_{i} \;\Bigl| \Bigr.\;
    \mathcal{P}\bigl(\frac{dN}{dt} \cdot t_i^\text{exp}\bigr) +
    \mu_{i}^{\text{bkg}}, \sigma_{i}^{\text{bkg}}\Bigr)
    \label{eq:likelihood}
\end{equation}
with the rate of reconverted photons $dN/dt$ as the signal parameter. The index $i$ iterates over the integer number of runs corresponding to a certain polarization state and $\mathcal{N}$ represents the Gaussian background parametrization of the $i$-th frame pair including an additional Poissonian signal contribution $\mathcal{P}\bigl(\frac{dN}{dt} \cdot t_i^\text{exp} \bigr)$ of expectation value $dN/dt$, multiplied by the frame pair exposure time $ t_i^\text{exp}=2 \times 5400$\,s. The numerical integration was performed via Markov-Chain-Monte-Carlos using twenty million samples for each photon polarization state, leading to a negligible uncertainty due to the integration process. 
The posterior probability distribution function  $p(dN/dt ~|~ Data)$ of reconverted photon rate $dN/dt$ deduced from the considered likelihood model assuming a flat prior distribution for the signal parameter $dN/dt$ are displayed Fig.~\ref{fig:postdistr} for the two linear polarizations of the incoming photon beam. 

\begin{figure}[t]
\begin{tabular}{c}
   \includegraphics[width=0.38\textwidth]{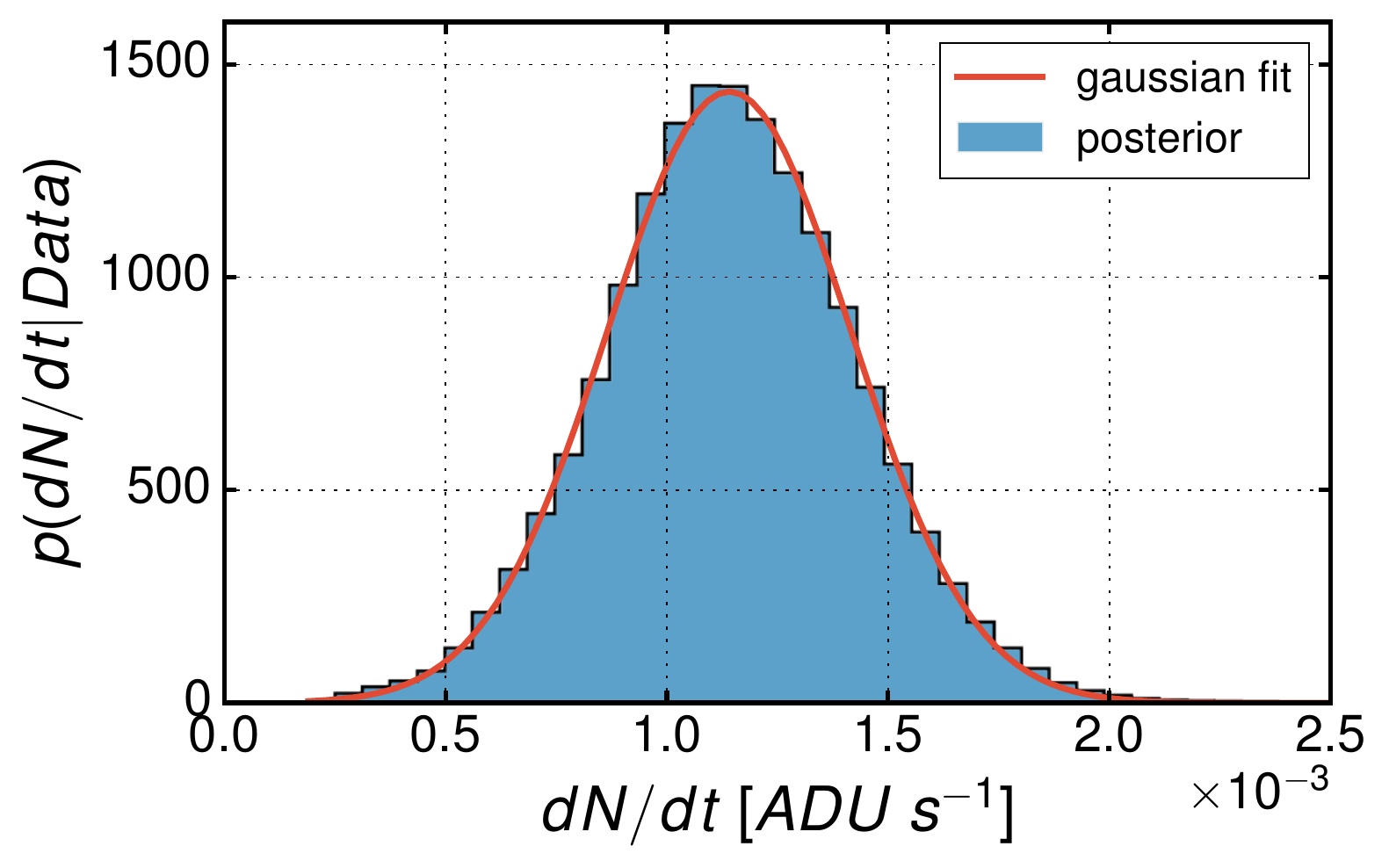}
 \\
   \includegraphics[width=0.38\textwidth]{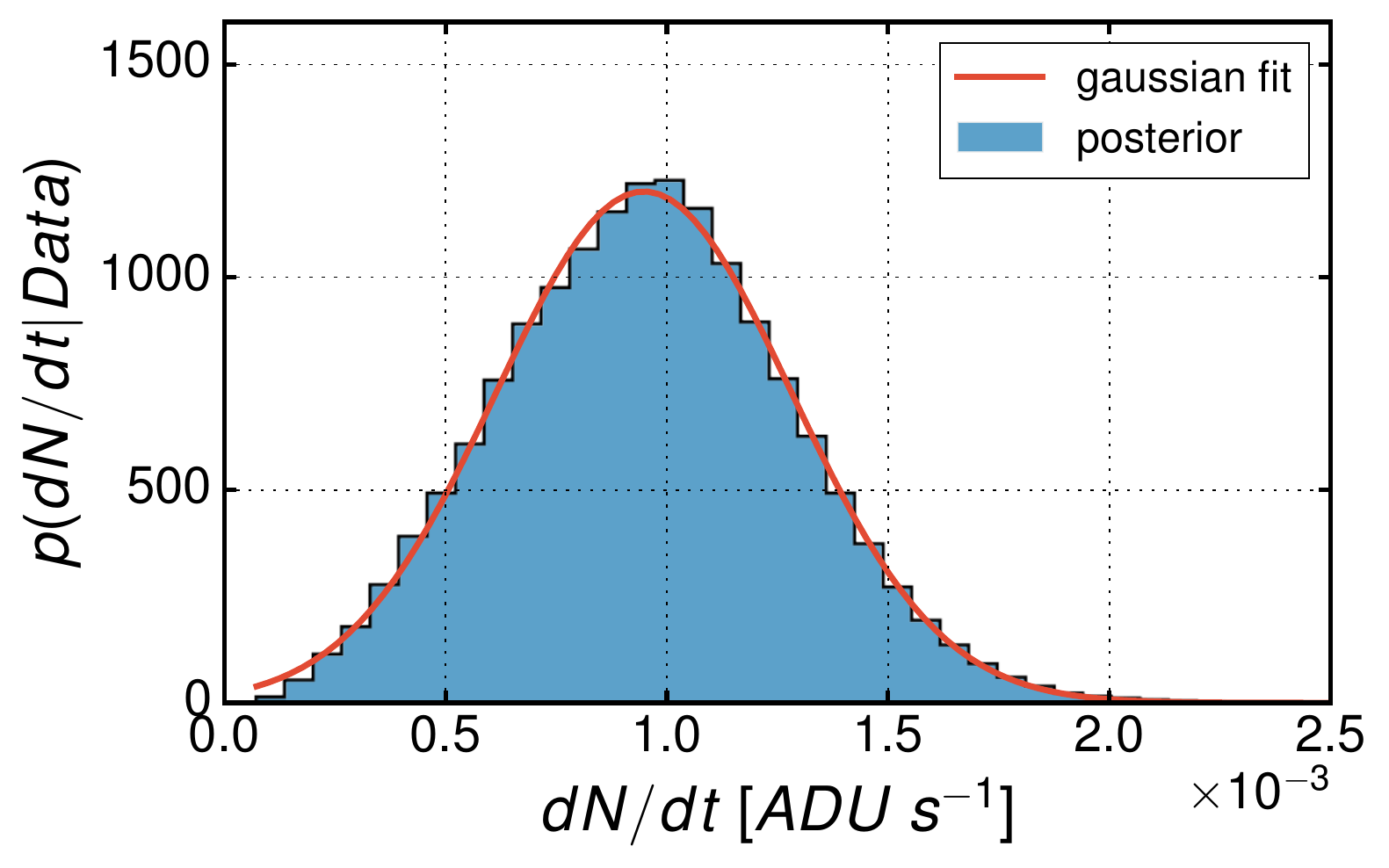}
\end{tabular}
    \caption{ \co Posterior probability distribution function  $p(dN/dt ~|~ Data)$ of reconverted photon rate $dN/dt$ for incoming photon polarization parallel ($\vec{E}_\gamma \parallel \vec{B}$) (Top) and perpendicular ($\vec{E}_\gamma \perp \vec{B}$) (Bottom) to the magnetic field $\vec{B}$, with an artificially imposed fake-signal of $1\times 10^{-3}$\,ADU\,s$^{-1}$.}
    \label{fig:fake}
\end{figure}

The consistency of the overall data analysis, including filtering and statistical methods was checked by imposing into all selected raw data frames fake-signals with a rate of 1\,mHz corresponding to an hypothetical ALP with a di-photon coupling constant $g_{\text{A}\gamma\gamma}= 3.8~10^{-8}$~GeV$^{-1}$ in the nearly massless limit ${(m_A \rightarrow 0)} $. 
A flux $dN/dt = 1.14\pm 0.28$\,mHz was inferred from the posterior probability distribution function  $p(dN/dt ~|~ Data)$ of reconverted photon rate $dN/dt$ for incoming photon beam linearly polarized parallel to the magnetic field and $dN/dt = 0.95\pm 0.33$\,mHz for incoming photon beam linearly polarized perpendicular to the magnetic field (Fig.~\ref{fig:fake}).

\section{Exclusion Limits}
\label{sec:exclusionlimits}

As a conclusion, no significant excess over the background expectation has been observed in the signal regions neither for the parallel nor the perpendicular polarization. Exclusion limits on the di-photon coupling strength $g_{\text{A}\gamma\gamma}$ and the ALPs mass $m_A$ have then been derived from Eqs.~\ref{eq:probability}~\&~\ref{eq:flux}. 

\begin{figure}[h]
    \centering
    \includegraphics[width=0.48\textwidth]{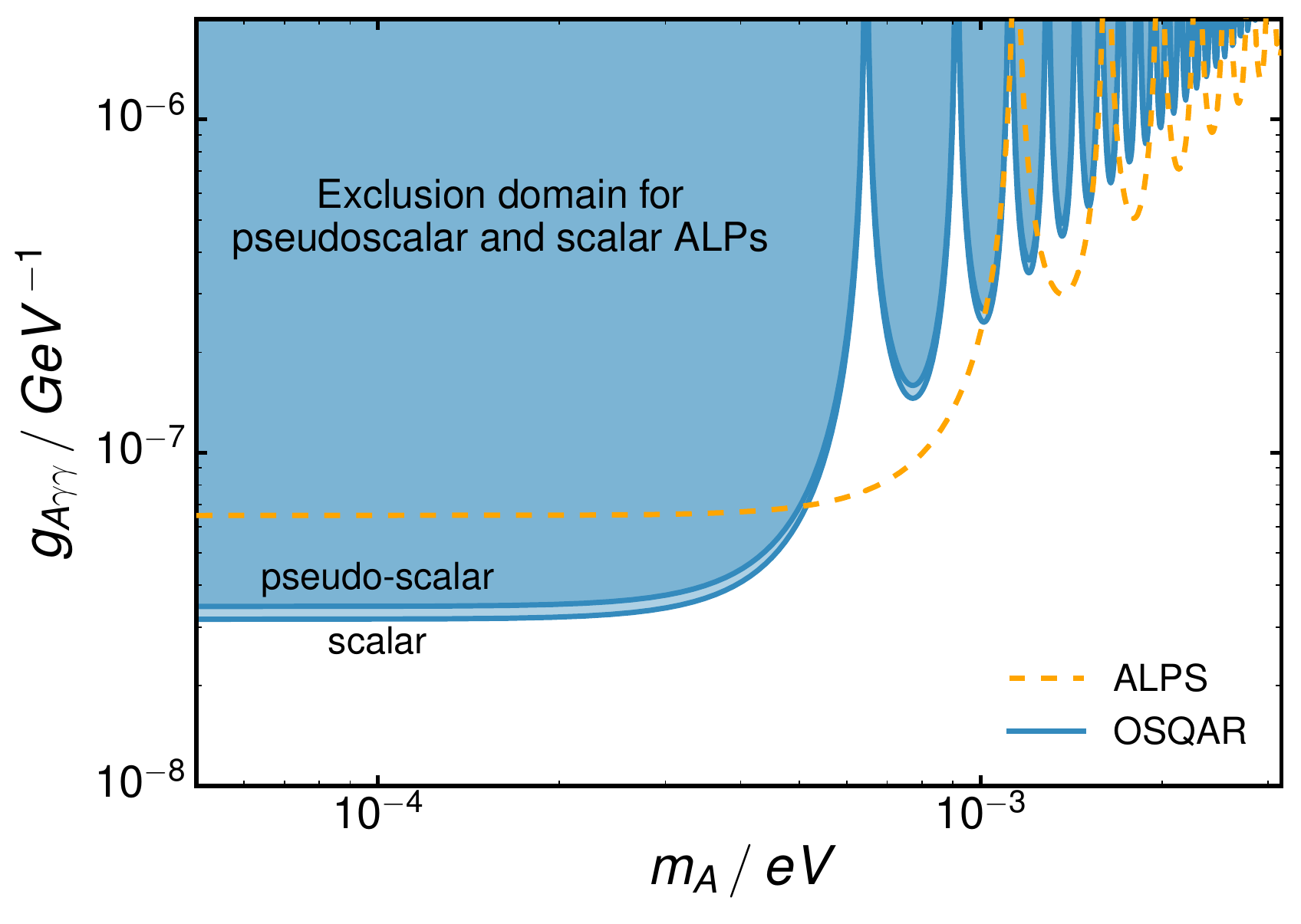}
    \caption{\co Exclusion limits for the searches of PS-ALPs and S-ALPs at 95\% C.L. obtained in vacuum by the present OSQAR LSW experiment. Also reported for comparison is the latest result of ALPS LSW experiment in vacuum.}
    \label{fig:limits}
\end{figure}

The $95\%$ Confidence Limit (C. L.) on the reconverted photon flux is derived via the posterior distribution of the signal parameter $dN/dt$ (Fig.~\ref{fig:postdistr}). The results for the PS-ALPs and S-ALPs searches are summarized in Fig.~\ref{fig:limits}, which also shows the exclusion limits reported by ALPS collaboration \cite{Ehret2010}. A limit for the di-photon couplings of $g_{\text{A}\gamma\gamma}<3.5\cdot10^{-8}$\,GeV$^{-1}$ and $g_{\text{A}\gamma\gamma}<3.2\cdot10^{-8}$\,GeV$^{-1}$ is obtained for the pseudoscalar and scalar searches for $m_A < 2\times 10^{-4}$\,eV, respectively. These results establish the most stringent constraints on ALPs  searches in the nearly massless limit obtained so far in LSW experiments.

\begin{acknowledgements}   

This work was partly supported by the Grant Agency of the Czech Republic 203/11/1546, by the SGS 2015 21053 Grant of Technical University of Liberec and by the Cluster of Excellence PRISMA at the university of Mainz.

\end{acknowledgements}

\end{document}